\documentclass[aps,prl,reprint, amsmath, amssymb,superscriptaddress]{revtex4-1}
 
\usepackage{bm}
\usepackage[retainorgcmds]{IEEEtrantools}
\usepackage{graphicx}
\usepackage{mathrsfs}
\usepackage{amsmath}
\usepackage{amsfonts}
\usepackage{amssymb}
\usepackage{color}
\usepackage{times,txfonts}
\usepackage{nicefrac}
\usepackage{ragged2e}
\usepackage{tikz}
\usepackage{bbm}
\usepackage{mathtools}
\usepackage[colorlinks=true,linkcolor=blue,urlcolor=blue,citecolor=blue,pdfusetitle]{hyperref}

\usepackage{blindtext}

\begin{document}

\title{Queued quantum collision models}
\date{\today}
\author{Guilherme Fiusa}
\affiliation{Department of Physics and Astronomy, University of Rochester, Rochester, New York 14627, USA}
\email{gfiusa@ur.rochester.edu}
\author{Gabriel T. Landi}
\affiliation{Department of Physics and Astronomy, University of Rochester, Rochester, New York 14627, USA}
\email{glandi@ur.rochester.edu}

\begin{abstract}
Collision models describe the sequential interactions of a system with independent ancillas. 
Motivated by recent advances in neutral atom arrays, in this Letter we investigate a model where the ancillas are governed by a classical controller that allows them to queue up while they wait for their turn to interact with the system. 
The ancillas can undergo individual open dynamics while they wait, which could cause them to, e.g., decohere. The system, which plays the role of the server in the queue, can also undergo its own open dynamics whenever it is idle. 
% We propose a general framework of queued collision models, that combine stroboscopic quantum dynamics with queueing theory in classical stochastic processes. 
We show that this framework greatly generalizes existing approaches for quantum collision models, recovering the deterministic and stochastic formulations in the appropriate limits. 
Next, we show how the {combination of queueing dynamics with quantum collisions introduces rich dynamical phenomena, including phase transitions and a sharp dependence on the queue statistics.}
% introduces nontrivial effects in the quantum collisions, that can lead to different phases in the system-ancilla response.
% This transition represents a trade-off between the waiting time dynamics of the ancillas, and the idle-time dynamics of the system. 
% We illustrate the idea with a model of coherence transfer under noisy waiting dynamics.

% a model where both the server and the queued ancillas are described by qubits, subject to either amplitude damping noise or a dephasing. We show how the queue parameters can predict whether or not a steady state is reached, and how the overall dynamics is sensitive to the expected behavior of the queue. Finally, we discuss how a phase transition in the queue has physical consequences that can be explored by studying time averages of the populations and coherences of the system.
\end{abstract}

%%%% TO DO %%%%%
% (1) Discuss the idea of neutral atom arrays, connecting it with further work, such as tha tango queue or the Barabasi model for priorities
% (2) Discuss the critical exponents of the phase transition, in hopes of obtaining some fractional value like 1/3, 2/5, etc
% (3) Rearrange the references, adding more experimental references as well as the neutral atom array ones 

\maketitle{}

{\bf \emph{Introduction.}}$-$Open quantum systems traditionally describe the interaction with a many-body reservoir, whose properties are not very controllable. 
However, advances in quantum-coherent platforms motivate the design of synthetic open dynamics, in which the environment is more structured. 
The canonical example is cavity QED~\cite{Gerry2005introductory, Walther2006}, in which the environment is a stream of atoms that are sent toward a cavity in an orderly fashion. 
% The dynamics is still open, but the environment now has a much finer degree of control. 
Other important examples include cascaded quantum systems~\cite{Carmichael1993, Gardiner1993, Stannigel2012}, squeezed baths~\cite{Ro_nagel2014, Niedenzu2018} and dispersive resonator couplings~\cite{Lee2023, Pechal2018, Hann2019, Chamberland2022}.
Recent experiments with neutral atom arrays~\cite{Bluvstein2022, Bluvstein2023, Ebadi2021,  Manetsch2024} have taken this a step further and demonstrated the efficient use of classical controllers, which can physically move qubits at will, in order to put them in contact so that they can interact. 
These results motivate {further studies on engineered open dynamics}.
% With the rapid development of quantum technologies in recent years, the necessity of accurate methods to study quantum systems becomes even more apparent.  Quantum systems are tricky entities, they are susceptible to errors and noise \cite{Preskill2018}; many-body systems are notoriously hard to study given their dimensionality; and contrary to classical systems, they display new types of correlations that can be thought of as resources for certain tasks \cite{Adesso2018, Adesso2017, Plenio2007, Horodecki2009, Gour2019, Gour2017, Brandao2015, Goold2016}. The realization of quantum systems in experiments such as ion traps, superconducting qubits, and ultracold atoms imply that those systems interact with an environment, so a faithful description of open dynamics is indispensable. The description of such dynamics requires certain approximations such as weak coupling with the bath and high Markovianity, and the description typically falls under quantum master equations that have a limited applicability scope. 

\begin{figure*}
    \centering
    \includegraphics[width=0.99\textwidth]{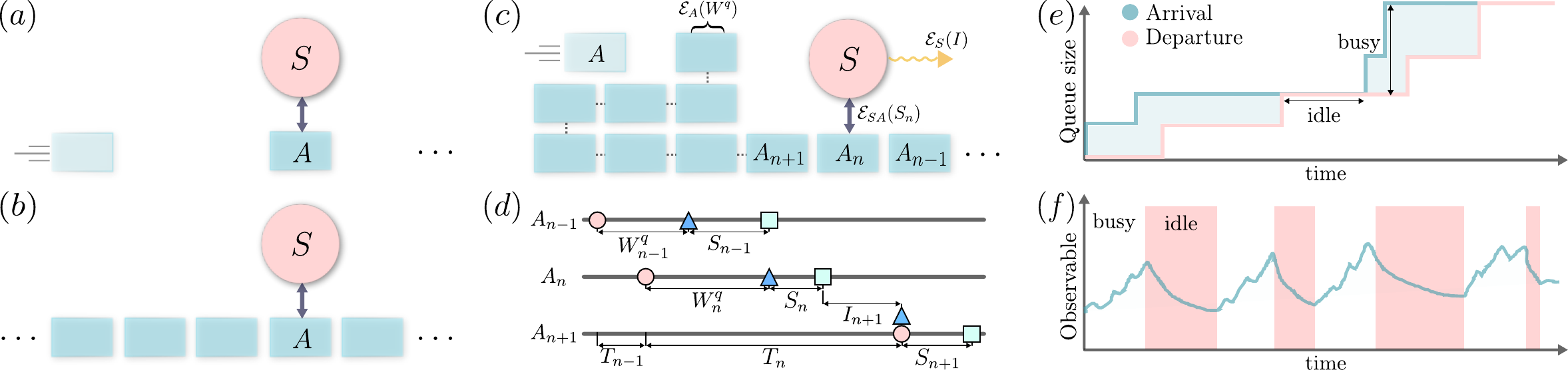}
    \caption{Quantum collision models (QCMs), where a system interacts sequentially with arriving ancillas. 
    $(a)$ Stochastic QCM, where ancillas arrive according to a random interarrival distribution.
    $(b)$ Deterministic QCM, where a ``conveyor belt'' of ancillas interacts stroboscopically with the system, at regular intervals. 
    $(c)$ The queued QCM introduced in this letter. 
    The system interacts with the ancilla via the quantum channel $\mathcal{E}_{SA}$.
    If an ancilla arrives and another one is already interacting with the system, it will queue up and wait for its turn. 
    While they wait, ancillas can undergo a quantum channel $\mathcal{E}_{A}$ that depends on the time spent in the queue $W_n^{q}$. 
    If no ancillas are waiting,  the system becomes idle and goes through a quantum channel $\mathcal{E}_{S}$ that depends on the idle period duration $I_n$. 
    $(d)$ Circuit diagram describing the main quantities in a classical queueing dynamics. Circles represent the ancilla's arrivals, triangles describe the beginning of the system-ancilla interaction and squares represent the ancilla's departures. 
    The waiting and idle times $W_n^q$ and $I_n$ are determined from the interarrival and service times $T_n$ and $S_n$ through the Lindley recursion relations \eqref{eq:lindley-recursion} and \eqref{eq:lindley-recursion-2}. 
    $(e)$ Schematic depiction of the random queue size over time, displaying busy and idle periods of the server (system). $(f)$ Schematic depiction of a system observable over time, showing how it responds differently to idle dynamics and system-ancilla interactions.
    }
    \label{fig:diagrams-and-timeline}
\end{figure*}

An approach that has enjoyed significant success in this respect are the so-called quantum collision models (QCMs)~\cite{Rau1963, Ziman2002, Ziman2005, Bruneau2014, MartinezMartinez2016, Strasberg2017, Campbell2021, Ciccarello2022, Lorenzo2017, De_Chiara2020, Taranto2020, Campbell2018, Chisholm2021, Cattaneo2022, Cattaneo2021, Guarnieri2020}, in which a central system interacts sequentially with arriving ancillas, one at a time (Fig.~\ref{fig:diagrams-and-timeline}(a,b)). 
QCMs replace the complex system+bath evolution with a simpler dynamics involving only two bodies, at any given time. 
QCMs allow for fine control over energetics and memory effects, making them interesting for the study of thermodynamics~\cite{DeChiara2018, Ziman2002,Melo2022,Rodrigues2019, Cusumano2018} and non-Markovianity~\cite{Massimo2017, Campbell2018, McCloskey2014, Kretschmer2016, Bernardes2014, Saha2024}.
For this reason, in the past they have been used in various tasks, such as
metrology/thermometry~\cite{Shu2020, Landi2019, Campbell2021, Tabanera2022, Strasberg2019, Barra2015, Cusumano2024}, modeling of continuous measurements~\cite{Gross2018, Belenchia2020, Landi2022}, renewal processes~\cite{Vacchini2016, Vacchini2020, Megier2021}, and quantum battery charging protocols~\cite{Seah2021, Salvia2023, Shaghaghi2022, Rodriguez2023}.

QCMs come in two standard flavors.
The first, called stochastic QCMs (Fig.~\ref{fig:diagrams-and-timeline}(a)), is motivated by Boltzmann's molecular chaos hypothesis.
It consists of a system that evolves unitarily by itself but is also subject to random collisions by arriving ancillas. 
The interarrival times of the ancillas are random, the duration of the collision process is short and, afterwards, the ancillas leave the process and never interact with the system again. 
The other flavor correspond to the so-called deterministic QCMs, in which there is a ``conveyor belt'' of ancillas, usually prepared in identical states (Fig.~\ref{fig:diagrams-and-timeline}(b)). 
Each ancilla interacts with the system for a fixed duration, after which they leave and never participate again in the dynamics. 
The reduced dynamics of the system is therefore stroboscopic. 
This kind of model fits naturally with cavity QED~\cite{Osnaghi2001, Gong2018, Maffei2022}, and also appears {in quantum optics, whenever the electromagnetic field is discretized in time bins}~\cite{Ciccarello2017,Gross2018}.

Both types of QCMs can be imagined as particular cases of a more general scenario, in which a classical controller 
physically moves the ancillas according to some protocol.  
{In the stochastic QCM of Fig.~\ref{fig:diagrams-and-timeline}(a) the controller randomly fires particles at the system; and in the deterministic QCM (Fig.~\ref{fig:diagrams-and-timeline}(b)) the controller aligns them perfectly, and ensure that they all interact with the system for the same duration.
While these are two natural choices, one could also consider more general controllers, which introduce new dynamical rules for how the system interacts with each ancilla.}
This will be the basic paradigm we adopt in this Letter: the system-ancilla interaction is quantum, but the way in which the ancillas are moved around is classical, governed by an external controller.
In neutral atom arrays, this paradigm is the backbone of any quantum computation~\cite{Bluvstein2022, Bluvstein2023, Ebadi2021}.
Our interest here will be in exploring the different types of dynamical behaviors that {emerge from different controller protocols.
There is already a plethora of new effects that emerge from comparing stochastic and deterministic QCMs. It is therefore natural to expect that new controller protocols should lead to rich new physics.
}

{To illustrate that, we consider the introduction of just a simple new ingredient. Namely, the assumption that when an ancilla arrives, it will queue up and wait for its turn (Fig.~\ref{fig:diagrams-and-timeline}(c)) if the system is occupied with another ancilla.}
%if the system is occupied with another one, that ancilla will queue up and wait for its turn (Fig.~\ref{fig:diagrams-and-timeline}(c)).}
% {a QCM with the following rules} (). 
% The system interacts with the ancillas one at a time, {for a duration that can be either random or deterministic.}
% Ancillas arrive according to some classical protocol, which can also be random or deterministic.
% The new ingredient {we introduce} is the assumption that, 
As we show, this leads to an incredibly rich set of possible behaviors, {since it allows one to introduce additional dynamics for whenever the system is idle (i.e., not interacting with ancilla), as well as for when the ancilla is waiting in the queue.}
In fact, through a simple qubit model, we illustrate how this competition between interaction and individual waiting/idle dynamics can lead {to effects such as steady state phase transitions and sharp dependence on the queue statistics.}

{\bf \emph{Queueing theory.}}$-$We consider a classical controller {which brings the ancillas sequentially, and queues them up while they wait for their turn to interact with the system. 
The system plays the role of a server, and the duration of their interaction is called the service time. After the interaction, the ancillas leave and do not participate again in the dynamics.} 
%This therefore forms a single server queuing process, with first come first serve discipline
The ancillas are labeled
as $n=1,2,3,\cdots$, based on the order in which they arrive.
We let $T_n$ denote the interarrival time between ancilla $n$ and $n+1$, with $T_1=0$ (i.e., we start counting from the moment that the first ancilla arrives). And we let $S_n$ denote the service/interaction time between system and ancilla $n$
(Fig.~\ref{fig:diagrams-and-timeline}(d)).
As we describe below, the set  $\{T_n,S_n\}$  completely specifies the classical queueing structure.
And, in principle, no additional assumptions need to be made about them:
they can be deterministic or stochastic, and they can be statistically independent or correlated.
In the simplest case (known as a G/G/1 queue) the $\{T_n,S_n\}$ are taken to form two iid sets, drawn from distributions $p_T(t)$ and $p_S(s)$.
{Let $\lambda = 1/E(T)$ denote the average interarrival rate, and $\mu = 1/E(S)$ the average service rate. The steady state properties of the queue are determined by the ratio 
\begin{equation}\label{rate}
    r := \frac{\lambda}{\mu}.
\end{equation}
If $r>1$ the queue builds up indefinitely, while if $r<1$ the queue population remains finite.}

{The unique feature of a queuing system is the interplay between idle and waiting times. If there are no ancillas in the queue immediately after a service is complete, the system becomes idle until the next ancilla arrives. 
In this case, the new ancilla does not have to wait and service starts immediately. 
On the other hand, if the system is busy, any ancilla that arrives will have to wait for some time. }

% {On top of the classical queue, our goal here is to introduce a quantum dynamics between system and ancillas. For standard collision models we only need to specify the system-ancilla interaction. But the queue structure opens new possibilities. In particular, there are two entirely new features. 
% The first is the existence of random periods where the system is idle (no ancillas to be served). 
% The second is that, while some ancillas have a finite waiting time in the queue, there is always a nonzero probability that they do not have to wait at all.
% In a queued collision model, we can therefore also include individual dynamics for the system, }

{The idle and waiting times can be determined from the set $\{T_n,S_n\}$, through Lindley's recursion relations~\cite{Lindley1952,Kleinrock1974queueing, Gross2011fundamentals, Cohen1969single}.}
We denote by $W_{n}^{q}$ the time the $n$-th ancilla waits in the queue~\footnote{The superscript ``q'' is to emphasize that this is the time spent in the queue, and not the overall time spent in the dynamics (queue + service) which would be $W_n = W_n^q +S_n$.}, and by $I_{n}$ the idle time of the system after the $(n-1)$-th ancilla leaves and before the $n$-th ancilla arrives. 
% Both of these quantities are non-negative, but can be zero.
{From the graphical representation in Fig.~\ref{fig:diagrams-and-timeline}(d) ones deduces that~\cite{Lindley1952}}
\begin{align}
\label{eq:lindley-recursion}
W_{n+1}^q &= \max\big\{0,
W_n^q + S_n - T_n\big\},
\\[0.2cm]
I_{n+1} &= \max\big\{0,
-(W_n^q + S_n - T_n)\big\}, 
\label{eq:lindley-recursion-2}
\end{align}
where $W_{1}^{q} \equiv 0$  (the first customer always finds the queue empty) and $I_{1} \equiv 0$. 
% These relations can be intuitively understood from the diagram in Fig.~\ref{fig:diagrams-and-timeline}(d).
One sees that $W_{n}^{q}I_{n}=0$, meaning either there is no waiting for the ancilla or no idleness for the system. 
That is, when the $n$-th ancilla arrives to find the system idle, it does not have to wait ($I_n \neq 0$ and $W_n^q=0$).
And if the ancilla $n$ is queued up when the system is done with ancilla $n-1$, the system will have no idle time ($I_n=0$ and $W_n^q \neq 0$).
{This property is further illustrated in Fig.~\ref{fig:diagrams-and-timeline}(e), in terms of the cumulative number of arrivals and departures.}
% Whenever their difference is nonzero, it means that the queue population is finite, and hence the system will be busy and the ancillas will have to wait. If they coincide, the queue is empty and the system is idle.} 
%See Supplemental Material \cite{supp} for a comprehensive discussion about the statistics of waiting times $W_n^q$ and idle times $I_n$.

{The absolute time at which the $n$-th ancilla arrives in the queue is $t_n = \sum_{j=1}^{n-1} T_j$, and the absolute time when it leaves is $s_n=t_n + W_n^q + S_n$ --- i.e., one adds to $t_n$ the time the ancilla spent waiting and the time it spent interacting with the system.}

%\begin{figure}
%    \centering
%    \includegraphics[width=0.45\textwidth]{diagram-timeline-arrow2.pdf}
%    \caption{Timeline for a typical scenario of the queue. Circles represent the arrival of ancillas, squares the departure of ancillas, and triangles the beginning of the service. Time intervals are explicitly displayed, so that the recursion relations \eqref{eq:lindley-recursion} and \eqref{eq:lindley-recursion-2} can be readily read. Each horizontal line denotes an ancilla, and the global time flows from left to right. In this particular example, the $(n-1)$-th and $n$-th ancillas arrive before the previous service ended, so they have to wait in the queue and the server has no idle time, $I_{n} = 0$. The $(n+1)$-th ancilla arrives after the service for the $n$-th was concluded, so the server was idle and there was no waiting time in the queue, $W_{n+1}^{q} = 0$.}
%    \label{fig:timeline}
%\end{figure}

{\bf \emph{Queued QCMs.}}$-$We assume, for simplicity, that all ancillas arrive in the queue prepared in the same state $\rho_A$. 
A closed-form stochastic equation can be obtained if we model the system dynamics in discrete steps after each ancilla \emph{leaves} the process. 
{The queued QCM is then fully specified by three quantum channels: (i) 
the dynamics of the ancilla while it waits, $\mathcal{E}_A(W_n^q)[\rho_A]$; 
(ii) the dynamics of the system while it is idle, $\mathcal{E}_S(I_n)[\rho_S]$; 
(iii) and the joint system-ancilla interaction $\mathcal{E}_{SA}(S_n)[\rho_S\otimes \rho_A]$.
The subscripts denote in which Hilbert spaces each channel acts, and the argument refers to the corresponding time.}
% While they wait, they may undergo a generic quantum channel $\mathcal{E}_A(W_n^q)[\rho_A]$ that depends on their waiting time $W_n^q$. 
% Similarly, if the system is idle, it may undergo its own quantum channel $\mathcal{E}_S(I_n)[\rho_S]$, which depends on the idle time $I_n$. 
%Both channels are assumed to reduce to the identity when $W_n^q=0$ or $I_n=0$. 
% If $W_n^q = 0$ or $I_n = 0$, their corresponding channels are assumed to reduce to the identity.
% Finally, the system-ancilla interaction is modeled by a channel \textcolor{black}{$\mathcal{E}_{SA}(S_n)[\rho_S\otimes \rho_A]$} that depends on the service time $S_n$ (see Fig.~\ref{fig:diagrams-and-timeline}(c)). 
% A pictorial representation of the response of the system to the fluctuating number of ancillas in the queue is shown in Fig.~\ref{fig:diagrams-and-timeline}(f). 
% During idle periods, the system evolves according to $\mathcal{E}_S$. During busy periods, it evolves according to a series of \textcolor{black}{$\mathcal{E}_{SA}$} maps (we assume that as soon as an ancilla leaves, the next starts interacting with the system immediately if there are ancillas in the queue).
% From a modeling perspective, it is more tractable to 
%Let $\rho_S^n = \rho_S(s_n)$ denote the state of the system after the $n$-th ancilla leaves (which occurs at absolute time $s_n$, Eq.~\eqref{eq:time_leaves_process}).
Denoting by $\rho_S^n$ the state of the system immediately after the $n$-th ancilla left (i.e. after the $n$-th collision), the stochastic map connecting $\rho_S^n$ with $\rho_S^{n-1}$ reads
\begin{equation}
\label{eq:reduced-state-overall-dynamics}
    \rho_{S}^{n} = \text{Tr}_{A}\left\{ \mathcal{E}_{SA}(S_{n})\left[ \mathcal{E}_{S}(I_{n})[\rho_{S}^{n-1}] \otimes \mathcal{E}_{A}(W_{n}^{q})[\rho_{A}] \right]\right\} ,
\end{equation}
where Tr$_{A}\{\cdot\}$ denotes the partial trace over the ancilla.
This is the map specifying the stochastic dynamics of the system.
The ensemble averaged (unconditional) evolution would be obtained by averaging over different realizations of the queue. 
In general, there is no closed deterministic equation for the unconditional evolution. 
Even though~\eqref{eq:reduced-state-overall-dynamics} is time-local, the unconditional dynamics obtained by averaging over multiple trajectories is generally non-Markovian~\cite{Ciccarello2013, Cattaneo2021}. 
% {However, in the Supplemental Material~\cite{supp} we show that such deterministic approach can be obtained, provided one works in an augmented space.}

Eq.~\eqref{eq:reduced-state-overall-dynamics} encompasses the stochastic and deterministic QCMs as particular cases: 
\begin{itemize}\itemsep-0.2cm
    \item The deterministic QCM of Fig.~\ref{fig:diagrams-and-timeline}(b) is recovered when all $S_n$ are equal (deterministic), $S_n \equiv \tau_{SA}$, and we set $\mathcal{E}_S = \mathcal{E}_A = \mathbb{I}$; i.e., system and ancilla are not affected by their idle/waiting dynamics. Eq.~\eqref{eq:reduced-state-overall-dynamics} then reduces to 
    \begin{equation}
        \rho_{S}^{n} = \text{Tr}_{A}\left\{ \mathcal{E}_{SA}(\tau_{SA})\left[\rho_{S}^{n-1} \otimes \rho_{A} \right]\right\},
    \end{equation}
    which is deterministic.
    \item The stochastic QCM of Fig.~\ref{fig:diagrams-and-timeline}(a) is recovered when $T_n \gg S_n$. The ancillas never have to wait, $W_n^q=0$ [Eq.~\eqref{eq:lindley-recursion}], so $\mathcal{E}_A = \mathbb{I}$. And the idle time equals the waiting time $I_n \simeq T_n$ [see Eq.~\eqref{eq:lindley-recursion-2}]. Eq.~\eqref{eq:reduced-state-overall-dynamics} becomes 
    \begin{equation}
        \rho_{S}^{n} = \text{Tr}_{A}\left\{ \mathcal{E}_{SA}(S_{n})\left[ \mathcal{E}_{S}(T_{n})[\rho_{S}^{n-1}] \otimes \rho_{A} \right]\right\}.
    \end{equation}
    The variables $T_n$ and $S_n$ are statistically independent, so it is possible to ensemble average the map, yielding
    % (see~\cite{supp} for details).
    % Ensemble averaging therefore yields 
    \begin{equation}
        \varrho_S^n = \text{Tr}_{A}\left\{ \bar{\mathcal{E}}_{SA}\left[ \bar{\mathcal{E}}_{S}[\varrho_{S}^{n-1}] \otimes \rho_{A} \right]\right\},
    \end{equation}
    where $\varrho_S^n= E(\rho_S^n)$ is the ensemble averaged (unconditional) state, 
    $\bar{\mathcal{E}}_{SA} = \int ds~\mathcal{E}_{SA}(s) p_S(s)$ is the system-ancilla channel averaged over the service time distribution $p_S(s)$, while $\bar{\mathcal{E}}_S = \int dt~\mathcal{E}_S(t) p_T(t)$ is the system idle dynamics, averaged over the interarrival times.    
\end{itemize}
\begin{figure}
    \centering
    \includegraphics[width=0.48\textwidth]{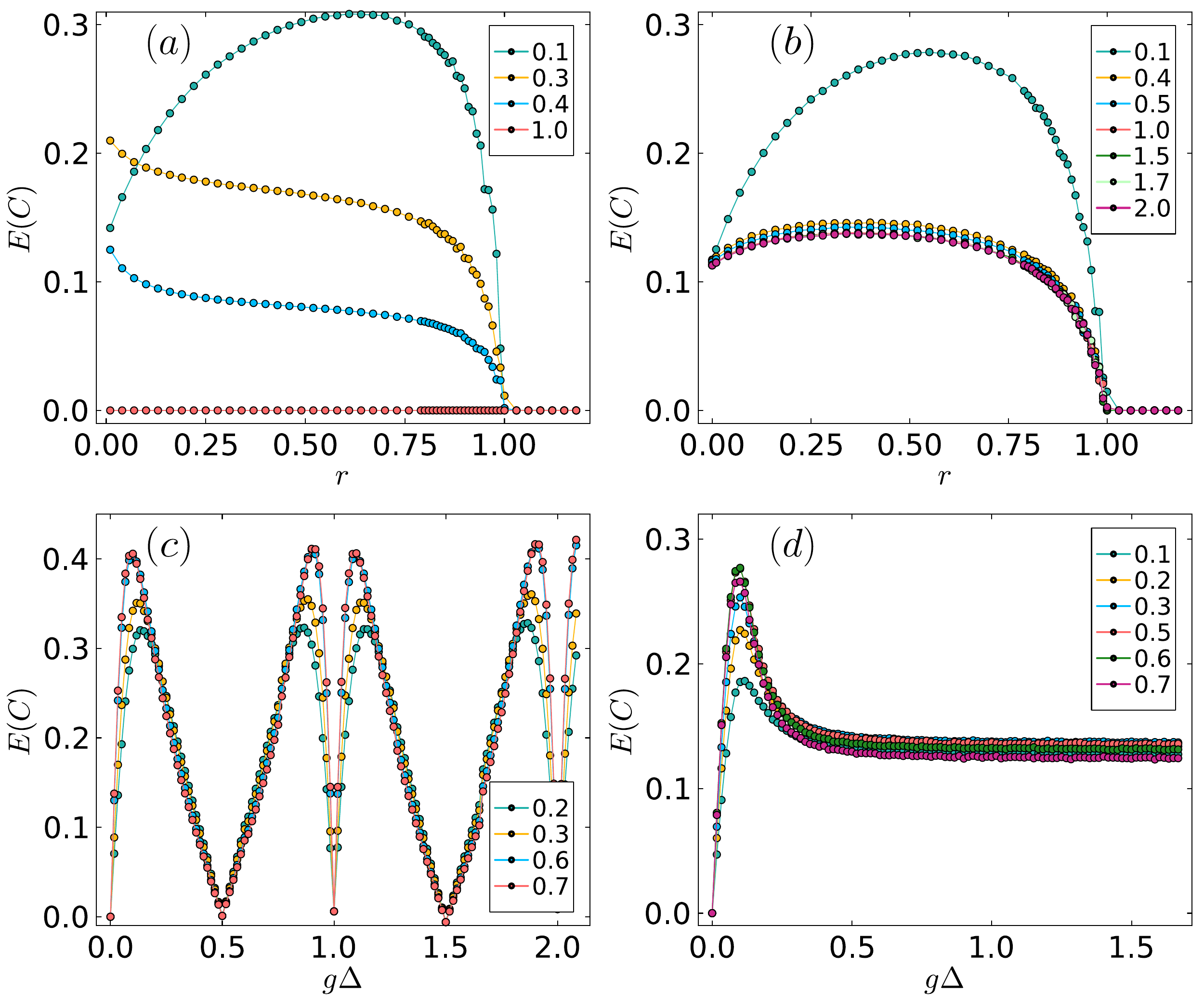}
    \caption{\textcolor{black}{Average of the coherence $E(\mathcal{C})$ in the long time limit as a function of $r$ for $(a), \ (b)$; $g\Delta$ for $(c), \ (d)$. For all plots we considered $g = \pi/12$, and $\gamma = 0.05$. Panels $(a)$ and $(c)$ consider a queueing process where the interarrival times $T_n$ are exponentially distributed with average $1/\lambda$ and service times are all constant given by $S_n = 1/\mu$, i.e. an M/D/1 queue. Different curves represent different values of $g\Delta$. Panels $(b)$ and $(d)$ consider a queueing process where both interarrival $T_n$ and service times $S_n$ are exponentially distributed, with averages $1/\lambda$ and $1/\mu$, respectively, characterizing an M/M/1 queue. Different curves represent different values of $r$.}
   % as a function of ancilla number (evaluated at the times each ancilla leaves the process), for different values of $\lambda \tau_{SA}$, with $\gamma\tau_{SA} = 0.05$ and $g\tau_{SA}=\pi/12$. (b) Single realization time average and (c) variance of the coherence in the long time limit as a function of $\lambda \tau_{SA}$, for different values of $g\tau_{SA}$, with $\gamma\tau_{SA} = 0.05$. 
    % $r$ and for different partial SWAP couplings $g$, in a singleshot regime. For $r>1$, the system reaches the steady state $\frac{1}{2}\mathbb{I}$ which is the state of the completely dephased ancillas. For $r<1$, the competition between the idle and waiting time dynamics gives rise to oscillations whose averages depend on the coupling strength $g$. For $r=1$, a phase transition takes place, irrespective of the coupling strength, which can be seen by the behavior of the average coherence going sharply to zero and the variance having a divergent behavior.
    }
    \label{fig:dynamics}
\end{figure}

%\begin{figure}
%    \centering
%    \includegraphics[width=0.48\textwidth]{dynplot_teste2.pdf}
%    \caption{Population and coherence of the system $S$ over ancilla interactions for the queued collision model. The system goes through an amplitude damping while it is idle and the ancillas dephase while they are waiting in the queue. System$-$ancilla interactions are partial SWAPs. In this realization, the physical parameters are $\gamma_{1} = 0.1$, $\gamma_{2} = 0.05$, $g=\pi/6$. For $r>1$, the system reaches the steady state $\frac{1}{2}\mathbb{I}$, which is the state of the completely dephased ancillas. For $r<1$, the competition between the idle and waiting time dynamics gives rise to oscillations in the population and coherence.}
%    \label{fig:dynamics}
%\end{figure}

{\bf \emph{Coherence evolution under an XXZ interaction.}}$-${The main goal in this Letter is to illustrate how the queueing structure can bring insights about new physics, that is not present in the stochastic or deterministic QCMs. To do that, we consider a model in which  both system and ancillas are qubits, interacting according to the XXZ interaction 
\begin{equation}\label{XXZ}
    H_{SA} = g(\sigma_x^S \sigma_x^A + \sigma_y^S \sigma_y^A + \Delta \sigma_z^S \sigma_z^A).
\end{equation}
All qubits are also subject to environmental dephasing, described by a quantum master equation with Lindblad dissipator $\mathcal{D}[\sigma_z](\rho)= \gamma (\sigma_z \rho \sigma_z-\rho)$.
The three channels $\mathcal{E}_S(I_n), \mathcal{E}_A(W_n^q)$ and $\mathcal{E}_{SA}(S_n)$ in Eq.~\eqref{eq:reduced-state-overall-dynamics} are thus specified in terms of the corresponding Liouvillians (semi-group generators) $\mathcal{L}_\alpha$ as
$\mathcal{E}_\alpha(t) = e^{\mathcal{L}_\alpha t}$ (with $\alpha=S,A,SA$). 
The idle/waiting dynamics are assumed to be given by pure dephasing, so $\mathcal{L}_\alpha = \mathcal{D}[\sigma_z^\alpha]$ for $\alpha=S,A$. 
For the system-ancilla interaction ($\alpha = SA$) we have pure dephasing in both system and ancilla, plus their interactions
\begin{equation}
    \mathcal{L}_{SA}(\rho_{SA}) = -i [H_{SA},\rho_{SA}] +  \mathcal{D}[\sigma_z^S](\rho_{SA})+\mathcal{D}[\sigma_z^A](\rho_{SA}). 
\end{equation}}
The ancillas are prepared in $\rho_A = |+\rangle\langle +|$,
where $|+\rangle = (|0\rangle + |1\rangle)/\sqrt{2}$.
We focus on how the coherence in the system, defined as $\mathcal{C} = |{\rm Tr}(\sigma_+\rho_S)|$~\cite{Baumgratz2014}, behaves in the steady state.
That is, we stochastically generate the queued collision model and average $\mathcal{C}$ over many collisions in a single run. 
If the ancillas wait for too long in the queue, they will decohere. 
And if the system is idle for very long periods, it will decohere as well.
However, as we will see, the competition between the effects can lead to nontrivial, and nonmonotonic behaviors.
% The dynamics of $\mathcal{C}$ will therefore look somewhat like Fig.~\ref{fig:diagrams-and-timeline}(f).
Examples of stochastic traces are presented in~\cite{supp}.
Here, we focus on the average $E(\mathcal{C})$ [Fig.~\ref{fig:dynamics}]. 
{We compare two queueing systems, the M/D/1 [Fig.~\ref{fig:dynamics}(a,c)] and M/M/1 [Fig.~\ref{fig:dynamics}(b,d)].
In both the interarrival time is given by a exponential distribution $p_T(t) = \lambda e^{-\lambda t}$, with arrival rate $\lambda$.
In the M/D/1 model we take the service time to be deterministic, $S_n = 1/\mu$. 
In the M/M/1 model we take it to also be exponentially distributed, $p_S(s) = \mu e^{-\mu s}$, with service rate $\mu$.
Figs.~\ref{fig:dynamics}(a,b) show results as a function of $r =\lambda/\mu$ [Eq.~\eqref{rate}], for different values of $g \Delta$. 
And Figs.~\ref{fig:dynamics}(c,d) show the results as a function of $g\Delta$ for different $r$.}

In Figs.~\ref{fig:dynamics}(a,b) we see a non-monotonic behavior of $E(\mathcal{C})$ with $r$, 
which reflects the nontrivial interplay of waiting and idle times. 
There is an optimal value $r\in [0,1]$, which depends on the quantum dynamics parameters $g$ and $\gamma$, for which the coherence reaches a maximum.
For $r = 1$ the behavior of $E(\mathcal{C})$ is nonanalytic. 
This happens because the queue population grows indefinitely for $r>1$, and therefore diverges in the steady state. 
As a consequence, the waiting times of the ancillas also diverge, and they completely decohere. 
In other words, for $r>1$ all ancillas interact with the system in the maximally mixed state $\rho_A = (1/2) \mathbb{I}$ (in the steady state). 
%In this case, the system reaches a steady state and this becomes a homogenization process.
%\textcolor{red}{For $r>1$ the system reaches a steady state. 
%But for $r<1$ it never does, in the sense that the long time dynamics} will cause the system state to fluctuate for arbitrarily long times, since the queue stochastically alternates between idle and busy periods, as depicted in Fig.~\ref{fig:diagrams-and-timeline}(f).
%This effect also becomes visible by plotting the variance of $\mathcal{C}$ in a single run (see~\cite{supp}).
In this case, the system reaches a true steady state. For $r<1$, the dynamics fluctuate for arbitrarily long times (alternating between idle and busy periods, see Fig.~\ref{fig:diagrams-and-timeline}(f)) so the system never reaches a steady state. This effect also becomes visible by plotting the variance of $\mathcal{C}$ in a single run (see~\cite{supp}).

{The behavior for $r<1$ in Figs.~\ref{fig:dynamics}(a,b) is quite different for the M/D/1 and M/M/1 models.
To better appreciate this, we  analyze the system as a function of $g\Delta$ for different $r$. 
The XXZ interaction~\eqref{XXZ} acts as a partial SWAP, that transfer part of the ancilla's coherence, but with a nontrivial phase depending on $\Delta$; the true partial SWAP is recovered only for $\Delta = 1$~\cite{Scarani2002}. 
In the M/D/1 [Fig.~\ref{fig:dynamics}(c)] we observe sharp oscillations of $E(\mathcal{C})$ as a function of $g\Delta$. 
And an almost perfect data collapse between curves for different $r$. 
Conversely, for the M/M/1 [Fig.~\ref{fig:dynamics}(d)] the behavior is markedly different: $E(\mathcal{C})$ has a peak at some finite $g\Delta$, and then saturates. The value at which it saturates also depends very weakly on $r$. 
This is a consequence of random phase averaging:  in the M/M/1 model each system-ancilla collision lasts for a different time, and therefore the phase picked up by the system randomly fluctuates as well. 
}

{\bf \emph{Conclusions.}}$-$We have introduced a new dynamical model of open quantum dynamics, in which a system interacts sequentially with ancillas, in a way that is governed by a classical controller. 
The only assumption is that the system interacts only with one ancilla at a time. 
All other dynamical properties are established by the choice of interarrival and service times $T_n$ and $S_n$, as well as the quantum maps $\mathcal{E}_{SA}$, $\mathcal{E}_A$ and $\mathcal{E}_S$.
This framework therefore generalizes collision models, containing previously studied QCMs as particular/limiting cases. 
Our goal with this Letter was to put forward this interesting connection between queued dynamics and collision models, and also to illustrate the new type of dynamics that emerges from such a connection. 
We did this through a minimal model of coherence transfer between qubits, \textcolor{black}{where we showed that the interplay between different queue statistics and dynamical parameters lead to nonanalytic behavior characteristic of phase transitions, as well as to strong dependences on the quantum model, which are not present in standard collision models.}

There is still much to be explored about the map~\eqref{eq:reduced-state-overall-dynamics}, including particular cases that might allow for analytical solutions, or other types of competing dynamics that could lead to interesting behaviors. 
In addition, the basic building blocks introduced here naturally lead to various other dynamical models.
First, one could allow the ancillas to interact with each other while in the queue. 
Second, one can introduce priority mechanisms, where certain ancillas are flagged as priorities, and therefore allowed to skip the queue entirely~\cite{Barabasi2005, Vazquez2005, Vazquez2006}.
%Third, one could study pairwise queues, where we have two queues and ancillas from each one interact pairwise with each other. 
Third, one could study pairwise queues, where ancillas from each queue interact with one another in a pairwise fashion, a setup normally referred to as ``taxicab'' queues.
This closely matches the quantum computations with neutral atom arrays~\cite{Bluvstein2023}. 
Fourth, one could introduce mechanisms in which the quantum dynamics also affect the classical queue. 
In Eq.~\eqref{eq:reduced-state-overall-dynamics} the classical queue dynamics affects the quantum properties, but not vice-versa. 
One way to change that is, for example, to have the system-ancilla service time be determined by the occurrence of a quantum jump in the system (e.g. a photon is emitted). 
{It is our hope that these future research directions motivate the study of synthetic open quantum dynamics, described by classical controllers.}

The authors acknowledge fruitful discussions with Gabriele De Chiara, Patrick Potts, Felix Binder, Felipe Pereira-Alves, Felipe Picoli, Giuliano Semente, and Abhaya Hedge.
% \section*{Acknowledgments}

\bibliography{local_library}

%\end{document}

\pagebreak
\widetext
%%%%%%%%%% Merge with supplemental materials %%%%%%%%%%

\newpage 
\begin{center}
\vskip0.5cm
{\Large Supplemental Material}
\end{center}
\vskip0.4cm

%%%%%%%%%% Prefix a "S" to all equations, figures, tables and reset the counter %%%%%%%%%%
\setcounter{section}{0}
\setcounter{equation}{0}
\setcounter{figure}{0}
\setcounter{table}{0}
\setcounter{page}{1}
\renewcommand{\theequation}{S\arabic{equation}}
\renewcommand{\thefigure}{S\arabic{figure}}

\section{Comments about queues and Derivation of Lindley's equation}
A queueing process is defined by the interarrival times $T_{n}$ and the service times $S_{n}$. Although the recursion relations \eqref{eq:lindley-recursion} and \eqref{eq:lindley-recursion-2} effectively provide a way to ``solve'' the queue by calculating the idle times $I_n$ and waiting times $W_n^q$ for every customer (ancilla), they make no mention toward the distributions of those quantities. Therefore, a natural question to address is what the distributions of $I_n$ and $W_n^q$ look like. One aspect that makes this analysis nontrivial and interesting is the fact that both distributions are a sum of a discrete part and a continuous smooth part, e.g., \begin{align*}
    P(W_n^q) &= P_{0}\delta (W_n^q) + f(W_n^q),\\
    P(I_n) &= I_{0} \delta (I_n) + g(I_n),
\end{align*}where $P_0$ and $I_0$ account for the nonvanishing probability that the ancilla has no waiting time (i.e. finds the queue empty upon arrival) or the system has no idle time, respectively. Functions $f$ ($g$) denote some smooth function that depends on $W_n^q$ ($I_n$). In what follows, we analyze some of the richness behind the distributions $P(W_n^q)$ and $P(I_n)$ in the case where both $T_n$ and $S_n$ are independent and identically distributed (iid) random variables.

Let us denote the cumulative distribution function (CDF) of the waiting times as
\begin{equation}
    F_{n}(x) = P\big(W_n^q \leqslant x\big).
\end{equation}
Here it is convenient to introduce $U_{n} = S_{n} - T_{n}$, which is an iid random variable distributed according to a probability density function $p(u)$. It follows,
\begin{align*}
F_{n+1}(x) &= P\big(W_{n+1}^q \leqslant x\big) \\
        &= P\big(W_{n+1}^q =0\big) + P\big(0< W_{n+1}^q \leqslant x\big)\\
&= P\big( W_n^q + U_n\leqslant 0\big ) + P\big( 0< W_n^q + U_n \leqslant x\big) \\
&= P\big(W_n^q + U_n \leqslant x\big)\\
&= \int_{-\infty}^{x} P\big(W_n^q \leqslant x-u) p(u) du,
\end{align*}where we conclude, after changing variables to $v = x-u$,
\begin{equation}
\label{eq:lindley-equation}
    F_{n+1}(x) = \int_{0}^{\infty}F_{n}(v)p(x-v)dv.
\end{equation}This result is known as the Lindley equation. It allows us to specify the CDF of any customer in terms of only the CDF of the previous customer. Given that $F_{n}(x)$ are CDFs, they have the following properties: (i) $F_{1}(x) = 1, \ \forall \ x \geqslant 0$, which follows from $W_{1}^{q} \equiv 0$; (ii) $F_{n}(x \to \infty) = 1$; (iii) $F_{n}(0)$ is generally not zero, since this represents the probability that the customer finds the queue empty and does not have to wait. 

Following the same procedure, we use the recursion relations \eqref{eq:lindley-recursion} and \eqref{eq:lindley-recursion-2} to derive a result where the idle time statistics can be calculated from the waiting time probabilities $F_{n}(x)$. The starting point is
\begin{equation*}
    P(I_{n+1} \leqslant x) = P(I_{n+1} = 0) + P(0 < I_{n+1} \leqslant x),
\end{equation*}applying the recursion relations \eqref{eq:lindley-recursion} and \eqref{eq:lindley-recursion-2},
\begin{align*}
    P(I_{n+1} \leqslant x) &= P(W_{n}^{q} + U_{n} = 0) + P(0 < -(W_{n}^{q} + U_{n} \leqslant x)\\
    &=P(W_{n}^{q} + U_{n}\geqslant 0) + P(-x \leqslant W_{n}^{q} + U_{n} < 0)\\
    &=P(-x \leqslant W_{n}^{q} + U_{n})\\
    &=P(W_{n}^{q} \geqslant -x - U_{n}) \ \ \ (x \geqslant 0),
\end{align*}now we use the fact that $U_{n}$ is an iid random variable with a PDF given by $p(u)$. It follows
\begin{align*}
     P(I_{n+1} \leqslant x) &= \int_{\infty}^{-\infty} du \ p(u) P(W_{n}^{q}  \geqslant -x-u)\\
     &= \int_{-\infty}^{\infty} du \ p(u) \left[1- P(W_{n}^{q}  \leqslant -x-u) \right]\\
     &=1-\int_{-\infty}^{-x}du \ p(u) F_{n}(-x-u),
\end{align*}so the result becomes
\begin{equation}
    G_{n+1}(x) = P(I_{n+1}  \leqslant  x) = 1 - \int_{0}^{\infty} dv \ p(-x-v) F_{n}(v),
\end{equation}with $v = -x-u$. To the best of the authors' knowledge, this result of queueing theory has never been reported in the literature before.

\section{Steady state solution for the deterministic qcm}
We further elaborate on the steady state solution of the deterministic QCM in Fig.~\ref{fig:diagrams-and-timeline}(b) from the queued QCM. Here, we consider $S_n \gg T_n$, so ancillas arrive much more quickly than the system can serve them. As a consequence, the queue quickly builds up, and the system will never be idle ($I_n = 0$), 
so $\mathcal{E}_{S}$ is the identity. To make a connection with typical deterministic QCM dynamics, we also assume that there is no waiting time dynamics for the ancillas. The dynamics from Eq.~\eqref{eq:reduced-state-overall-dynamics} then reduce to 
\begin{equation}
\label{eq:deterministic-col-model-appendix}
    \rho_{S}^{n} = \text{Tr}_{A}\left[\mathcal{E}_{SA}(S_{n}) (\rho_{S}^{n-1} \otimes \rho_{A})  \right].
\end{equation}The expectation value of the state after the collision with the $n$-th ancilla is then given by
\begin{align*}
    E[\rho_{S}^{n}] &= \int dS_{1}...dS_{n}P(S_{1})...P(S_{n}) \ \text{Tr}_{A}[\mathcal{E}_{SA}(S_{n})(\rho_{S}^{n-1}\otimes \rho_{A})]\\
&= \int dS_{n}P(S_{n}) \ \text{Tr}_{A}[\mathcal{E}_{SA}(S_{n})(E[\rho_{S}^{n-1}]\otimes \rho_{A})],    
\end{align*}and the steady state follows
\begin{align}\label{integration-steady state-deterministic}
    E[\rho_{S}^{ss}] &= \int ds  P(s) \ \text{Tr}_{A}[\mathcal{E}_{SA}(s)(E[\rho_{S}^{ss}]\otimes \rho_{A})]\\
    &= \Phi[E(\rho_{S}^{ss})].
\end{align}Note that the steady state solution is reached irrespective of the properties of the service times $S_{n}$. However, to recover the standard formulation, we may assume that $S_{n} = \tau_{SA}$, and that the system-ancilla map is unitary, of the form  $\mathcal{E}_{SA} \equiv \mathcal{U}_{SA}(\tau_{SA})[\bullet] = e^{-i H_{SA} \tau_{SA}} \bullet e^{i H_{SA} \tau_{SA}} $, where $H_{SA}$ is the system-ancilla Hamiltonian. We emphasize, however, that the steady state solution does not rely on this last assumption. In the case that $S_{n} = \tau_{SA}$, the integration~\eqref{integration-steady state-deterministic} becomes trivial.

\section{Recovering the stochastic QCM from the queued QCM}
{While the queued QCM dynamics is highly non-Markovian, the stroboscopic realization we described in the main text is Markovian by construction [Eq.~\eqref{eq:reduced-state-overall-dynamics}], because there we take into account discrete times taken when the service ends. This is known as a \textit{Markovian embedding}}. Here, we expand on this discussion and show how one recovers the stochastic model in Fig.~\ref{fig:diagrams-and-timeline}(a) from the queued QCM \eqref{eq:reduced-state-overall-dynamics}, considering a ``time ensemble''. We consider $T_n \gg S_n$, this implies, following Eq.~\eqref{eq:lindley-recursion-2}, that $W_{n+1}^q \simeq 0$ and $I_{n+1} \simeq T_{n}$, which effectively ``decouples'' the distributions of $T_n$ and $S_n$ with regards to their interplay for computing $I_n$ and $W_n^q$. The queue is empty most of the time, so ancillas arriving do not have to wait. The system is evolving, most of the time, under the idle map $\mathcal{E}_S(T_{n-1})$, which now depends only on the interarrival times $T_{n-1}$.
We also assume that $\mathcal{E}_{SA}(S_n) = \mathcal{E}_{SA}$ is still nontrivial, even for very small $S_n$. Under these assumptions, Eq.~\eqref{eq:reduced-state-overall-dynamics} reduces to
\begin{equation}\label{eq:stochastic-col-model_step0}
    \rho_S^n =\text{Tr}_{A}\left\{ \mathcal{E}_{SA}\left[ \mathcal{E}_{S}(T_{n-1})[\rho_{S}^{n-1}] \otimes \rho_{A} \right]\right\}.
\end{equation}
%while Eq.~\eqref{eq:time_leaves_process} reduces to $s_n = \sum_{j=1}^n T_j$.

It is clear that in this limit any stochasticity in the model will reside in the statistics of $T_{n-1}$. 
The usual stochastic QCMs correspond to the case where the interarrival times $T_n$ are iid with distribution $p_T(t)$.
A more common representation of stochastic QCMs is obtained if one switches to a time ensemble, where we describe the state of the system at a definite time, but allow the number of collisions to vary. 
This can be accomplished by defining
\begin{equation}
\label{conditional-state}
    \varrho_{S}^{(n)}(t) := E[\rho_{S}^{n} \delta(t-s_{n-1})]/E[\delta(t-s_{n-1})],
\end{equation}
where the average is over the joint realization of all interarrival times, and $s_n$ denotes the absolute time, given by $s_n:=\sum_{j=1}^n (I_j + S_j)$. The state $ \varrho_{S}^{(n)}(t)$ now describes the evolution of the system over some time $t$, where the number of collisions $n$ can vary. Explicitly, the expectation values are given by
\begin{align*}
    E[\cdot] &= \int dT_{1}...dT_{n-1}(\cdot) P(T_{1})...P(T_{n-1})\\[0.5em] 
   &\equiv \int dT_{1...n-1}(\cdot) P(T_{1...n-1}),
\end{align*}it then follows 
\begin{align*}
      \varrho_{S}^{(n)}(t) = \dfrac{1}{E[\delta(t-s_{n-1})]} \int dT_{1...n-1}P(T_{1...n-1})\delta(t-s_{n-1})\text{Tr}_{A}\left[\mathcal{E}_{SA}\left(\mathcal{E}_{S}(T_{n-1})(\rho_{S}^{n-1}) \otimes\rho_{A} \right)\right], 
\end{align*}where the only term that depends on the interarrival time $T_{n-1}$ is the idle dynamics channel. On top of that, we can substitute $s_{n-1} = s_{n-2} + I_{n-1} + S_{n-1} = s_{n-2} + T_{n-2}$ in the delta inside the integral, so we get
%\begin{widetext}
\begin{align*}
     \varrho_{S}^{(n)}(t) &= \dfrac{1}{E[\delta(t-s_{n-1})]} \int dT_{n-1} P(T_{n-1})\text{Tr}_{A}\big[\mathcal{E}_{SA}\left(\mathcal{E}_{S}(T_{n-1})\right. \left(\int P(T_{1...n-2})\rho_{S}^{n-1}\delta(t-s_{n-2}-T_{n-2})dT_{1...n-2}\right)\otimes \rho_{A}\big)\big],\\[0.5em]
     &= \dfrac{E[\delta(t-s_{n-2})]}{E[\delta(t-s_{n-1})]}\int dT_{n-1} P(T_{n-1}) \text{Tr}_{A}\left[\mathcal{E}_{SA}\left(\mathcal{E}_{S}(T_{n-1})[\rho_{S}^{(n-1)}(t-T_{n-1})]\otimes \rho_{A}\right)\right].
\end{align*}By defining $\Tilde{\varrho}_{S}^{(n)}(t) := \varrho_{S}^{(n)}(t)E[\delta(t-s_{n-1})]$ and changing variables $T_{n-1} \equiv t'$, we obtain the final result
\begin{equation}
\label{eq:stochastic-col-model2}
    \Tilde{\varrho}_{S}^{(n)}(t) = \int dt' p_{T}(t') \text{Tr}_{A}\left[\mathcal{E}_{SA}\left(\mathcal{E}_{S}(t')[\Tilde{\varrho}_{S}^{(n-1)}(t-t')]\otimes \rho_{A}\right) \right],
\end{equation}where $P(t') \equiv p_{T}(t')$. Eq. \eqref{eq:stochastic-col-model2} describes precisely the familiar representation used in studies of stochastic QCMs~\cite{Strasberg2017}.

\section{Variance of the coherence evolution under an XXZ interaction}

\begin{figure}
    \centering
    \includegraphics[width=0.48\textwidth]{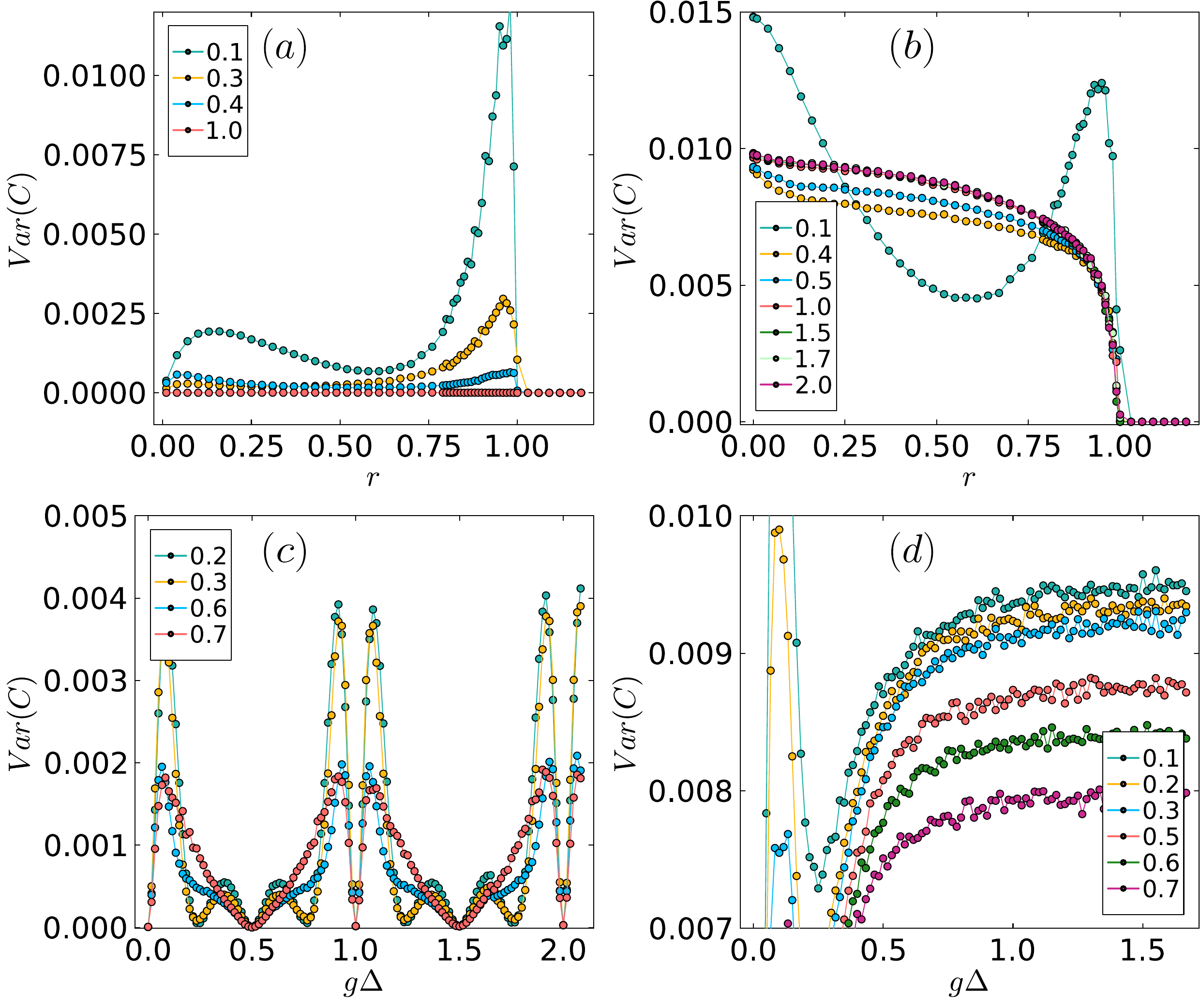}
    \caption{\textcolor{black}{Variance of the coherence $E(\mathcal{C})$ in the long time limit as a function of $r$ for $(a), \ (b)$; $g\Delta$ for $(c), \ (d)$. For all plots we considered $g = \pi/12$, and $\gamma = 0.05$. Panels $(a)$ and $(c)$ consider a queueing process where the interarrival times $T_n$ are exponentially distributed with average $1/\lambda$ and service times are all constant given by $S_n = \tau_{SA}$, i.e. an M/D/1 queue. Different curves represent different values of $g\Delta$. Panels $(b)$ and $(d)$ consider a queueing process where both interarrival $T_n$ and service times $S_n$ are exponentially distributed, with averages $1/\lambda$ and $1/\mu$, respectively, characterizing an M/M/1 queue. Different curves represent different values of $r$.}
   % as a function of ancilla number (evaluated at the times each ancilla leaves the process), for different values of $\lambda \tau_{SA}$, with $\gamma\tau_{SA} = 0.05$ and $g\tau_{SA}=\pi/12$. (b) Single realization time average and (c) variance of the coherence in the long time limit as a function of $\lambda \tau_{SA}$, for different values of $g\tau_{SA}$, with $\gamma\tau_{SA} = 0.05$. 
    % $r$ and for different partial SWAP couplings $g$, in a singleshot regime. For $r>1$, the system reaches the steady state $\frac{1}{2}\mathbb{I}$ which is the state of the completely dephased ancillas. For $r<1$, the competition between the idle and waiting time dynamics gives rise to oscillations whose averages depend on the coupling strength $g$. For $r=1$, a phase transition takes place, irrespective of the coupling strength, which can be seen by the behavior of the average coherence going sharply to zero and the variance having a divergent behavior.
    }
    \label{fig:dynamics-appendix-variance}
\end{figure}

In the main text, we discussed some features that appeared due to the queueing mechanism in the dynamics by studying the average coherence of the qubit model under an XXZ interaction. Here, we extend this discussion by exploring the variance of the coherence, $Var(\mathcal{C}) = E(\mathcal{C}^2) - E(\mathcal{C})^2$. In Fig.~\ref{fig:dynamics-appendix-variance} we plot the variance of the coherence as a function of $r$ and $g\Delta$ for the M/D/1 and M/M/1 queue models.

The behavior for $r<1$ in Figs.~\ref{fig:dynamics-appendix-variance}(a,b) is once again quite different for the M/D/1 and M/M/1 models. First, we see that in the M/D/1 case [Fig.~\ref{fig:dynamics-appendix-variance}(a)] the fluctuations are higher for small but nonzero $r$ as well as for $r\to 1$. In particular, we observe that for $r=1$ the fluctuations tend to diverge, which is a typical feature of phase transitions. In the M/M/1 case, we have different behaviors depending on the value of $g\Delta$. For the smallest one ($g\Delta = 0.1$), the behavior is similar to the one from the M/D/1 case where we observe peaks around $r\to 0$ and $r\to 1$. But as we increase the anisotropy, the fluctuations become monotonically decreasing, and in particular, all curves match once we reach $g\Delta \sim 1$. In both queueing models, the variance also becomes nonanalytic at $r=1$, irrespective of the $g\Delta$ value.

In Figs.~\ref{fig:dynamics-appendix-variance}(c,d) we explore how the fluctuations behave as a function of $g \Delta$, for different values of $r$. For the M/D/1 case [Fig.~\ref{fig:dynamics-appendix-variance}(c)], we observe sharp oscillations, but with a more irregular pattern in comparison with the oscillations observed for the average [Fig.~\ref{fig:dynamics}(c)]. Likewise, the behavior for the M/M/1 queue [Fig.~\ref{fig:dynamics-appendix-variance}(d)] also follows what we observed before [see Fig.~\ref{fig:dynamics}(d)], there is a peak at some finite $g\Delta$ followed by a saturation that depends weakly on the values of $r$. This suggests that the random phases that arise as a consequence of the anisotropy do not play a key role in the fluctuations.

\section{Further examples of the queued dynamics}
In the main text, we considered a qubit model of coherence transfer where both the system and the ancillas underwent a dephasing channel during their idle and waiting dynamics, respectively. \textcolor{black}{Their interaction was given by a quantum master equation that involved a partial SWAP as well as a dephasing, described by the jump operators.} The goal was \textcolor{black}{to prepare ancillas with coherence and observe the coherence transfer to the system through the noisy partial SWAP interactions as a function of $r$ [Eq.~\eqref{rate}] and $g\Delta$. We observed that the queueing process played an important role in the dynamics, in particular, we saw that the anisotropy $\Delta$ could severely enhance or suppress the transfer. However, one feature that deserves further attention is the competition between the different dynamical channels that arise as a consequence of the interplay of waiting/idle times. This is what we explore here.}

To that end, we consider a \textcolor{black}{simpler} model  where the system and ancillas are qubits, which exchange excitations via a perfect partial SWAP unitary 
$\mathcal{E}_{SA} \equiv \mathcal{U}_{SA}(S_n)[\rho_{S}\otimes \rho_{A}] = U_{S_n} (\rho_S \otimes \rho_A)U_{S_n}^\dagger$, where $U_{S_n} = i \cos(gS_n) + \sin(gS_n) U_{\tt swap}$.
Here $g$ controls the strength of the interaction and $U_{\tt swap}$ is the full SWAP (one can think that this interaction is an approximation where the dephasing during interactions is negligible).
% $U_{\tt swap} = \frac{1}{2}(1 + \sigma_x^S \sigma_x^A + \sigma_y^S \sigma_y^A + \sigma_z^S \sigma_z^A)$ 
% (here $\sigma_\alpha$ are Pauli matrices).
% $\mathcal{U}_{SA}(S)[\rho_{S}\otimes \rho_{A}] = e^{i S H_{SA}}(\rho_{S}\otimes \rho_{A})e^{-i S H_{SA}}$, where $H_{SA} = \frac{g}{2}(\sigma_{+}^{S}\sigma_{-}^{A} + \sigma_{-}^{S}\sigma_{+}^{A})$  
The ancillas are prepared in $\rho_A = |+\rangle\langle +|$, 
% which is the positive eigenstate of the Pauli matrix $\sigma_x$/
where $|+\rangle = (|0\rangle + |1\rangle)/\sqrt{2}$ and $|0/1\rangle$ are the eigenstates of the Pauli matrix $\sigma_z$.
The partial SWAP dynamics therefore transfers some of the coherence from the ancillas to the system. 
\textcolor{black}{We consider that the idle channel and the waiting time channel are both described by a dephasing channel
$\mathcal{E}(t)[\rho] = \frac{1}{2}\big[(1+e^{-\gamma t}) \rho + (1 - e^{-\gamma t})\sigma_z \rho \sigma_z\big]$ with $t \to I_n$ for the system and $t \to W_n^q$ for the ancillas.} 
The goal now is to observe the competition that maintains the system in a state with a high amount of coherence, which we quantify through the average $E(\mathcal{C}) = |\text{Tr}(\sigma_+ \rho_S)|$.
In this regard, the regimes where $r>1$ or $T_n \gg S_n$ are both deleterious. 
In the former, the ancillas wait too long and hence lose the coherence before they transfer it to the system.
In the latter, the system is idle too often and hence loses the coherences it receives from the ancillas. 
To illustrate this interplay more concretely, we assume that the service times are all equal and deterministic, $S_n = \tau_{SA} \equiv 1/\mu$, while the interarrival times $T_{n}$ are iid and exponentially distributed, with $p_T(t) = \lambda e^{-\lambda t}$, where $\lambda$ determines the rate of arrivals. \textcolor{black}{That is, we consider here an M/D/1 queueing process.}
The queueing properties are then fully determined by $r = \lambda/\mu$.
The quantum dynamics, on the other hand, is described by the interplay between the time-scales $g \tau_{SA}$, $\gamma I_n$ and $\gamma W_n^q$. 

Now, to \textcolor{black}{elucidate} the effect of each dynamical quantum channel separately (and then their competition), we consider two particular cases.
In the first, we consider that there is dephasing only in the system, so the idle channel is $\mathcal{E}_{S}(t)[\rho] = \frac{1}{2}\big[(1+e^{-\gamma t}) \rho + (1 - e^{-\gamma t})\sigma_z \rho \sigma_z\big]$ with $t \to I_n$, and the waiting time channel is the identity ($\mathcal{E}_{A} = \mathbb{I}$). In the second, we have a dephasing channel only for the ancillas, so the idle channel is the identity ($\mathcal{E}_S = \mathbb{I}$) and the waiting time channel is $\mathcal{E}_{A}(t)[\rho] = \frac{1}{2}\big[(1+e^{-\gamma t}) \rho + (1 - e^{-\gamma t})\sigma_z \rho \sigma_z\big]$ with $t \to W_n^q$.

\begin{figure}
    \centering
    \includegraphics[width=0.49\textwidth]{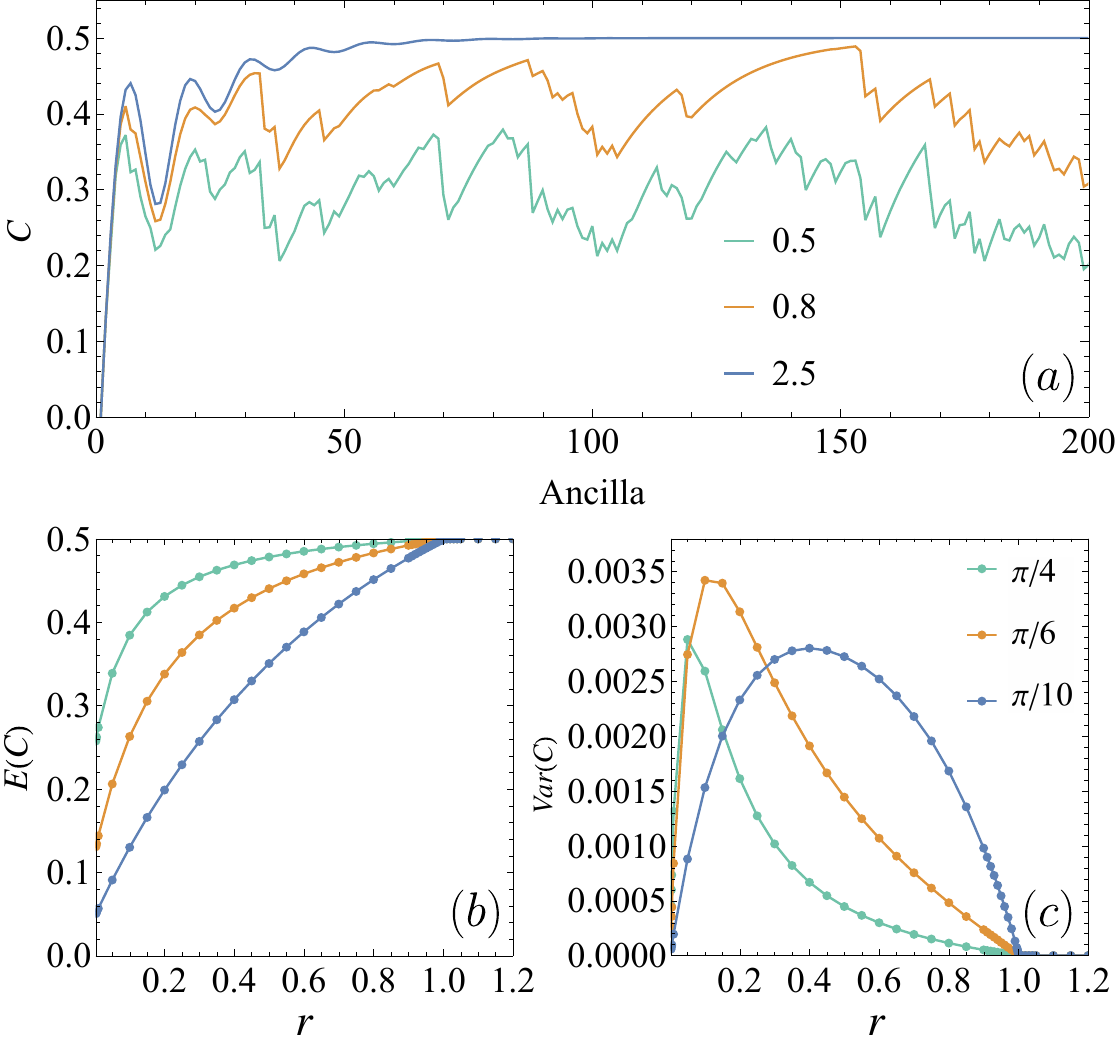}
    \caption{$(a)$ Single shot coherence $\mathcal{C}$ as a function of ancilla number (evaluated at the times each ancilla leaves the process), for different values of $r$, with $\gamma = 0.05$ and $g =\pi/12$. Here we consider only idle dephasing dynamics. $(b)$ Average coherence $E(\mathcal{C})$ and $(c)$ variance of the coherence $Var(\mathcal{C})$ in the long time limit as a function of $r$, for different values of coupling strength $g$, with $\gamma = 0.05$. 
    % $r$ and for different partial SWAP couplings $g$, in a singleshot regime. For $r>1$, the system reaches the steady state $\frac{1}{2}\mathbb{I}$ which is the state of the completely dephased ancillas. For $r<1$, the competition between the idle and waiting time dynamics gives rise to oscillations whose averages depend on the coupling strength $g$. For $r=1$, a phase transition takes place, irrespective of the coupling strength, which can be seen by the behavior of the average coherence going sharply to zero and the variance having a divergent behavior.
    }
    \label{fig:dynamics-dephasing-system}
\end{figure}

Let us begin by discussing the first \textcolor{black}{case}, where the dephasing channel describes only the idle system dynamics. In Fig.~\ref{fig:dynamics-dephasing-system}(a) we plot the coherence after each ancilla collision for different values of $r$. In this setup, ancillas always have coherence $1/2$ \textcolor{black}{(they are prepared in the state $\rho_A = |+\rangle\langle +|)$} whereas the system loses coherence whenever it becomes idle. This competition gives rise to the oscillations observed in Fig.~\ref{fig:dynamics-dephasing-system}(a) for $r < 1$.
Since there is no other mechanism in which the system may obtain coherence, we observe that the average coherence is monotonically increasing with respect to $r$, see Fig.~\ref{fig:dynamics-dephasing-system}(b). This result is intuitive in the sense that if the arrivals are more frequent, the system is less and less idle, hence it loses less and less coherence. Once $r > 1$ is reached, a transition takes place and the system becomes permanently busy. The process becomes a homogenization problem, where \[ \lim_{n \to \infty} \rho_S^n = \rho_{A}, \]
and a true steady state is reached because the fluctuations go to zero, see Fig.~\ref{fig:dynamics-dephasing-system}(c).
%This model captures the small $\lambda \tau_{SA}$ behavior (``left'' part of Fig.~\ref{fig:dynamics}(b)) of the general qubit model presented for the QQCM, where we observe that $\mathcal{C}$ is monotonically increasing with $\lambda \tau_{SA}$ before the inflection point is reached.

\begin{figure}
    \centering
    \includegraphics[width=0.49\textwidth]{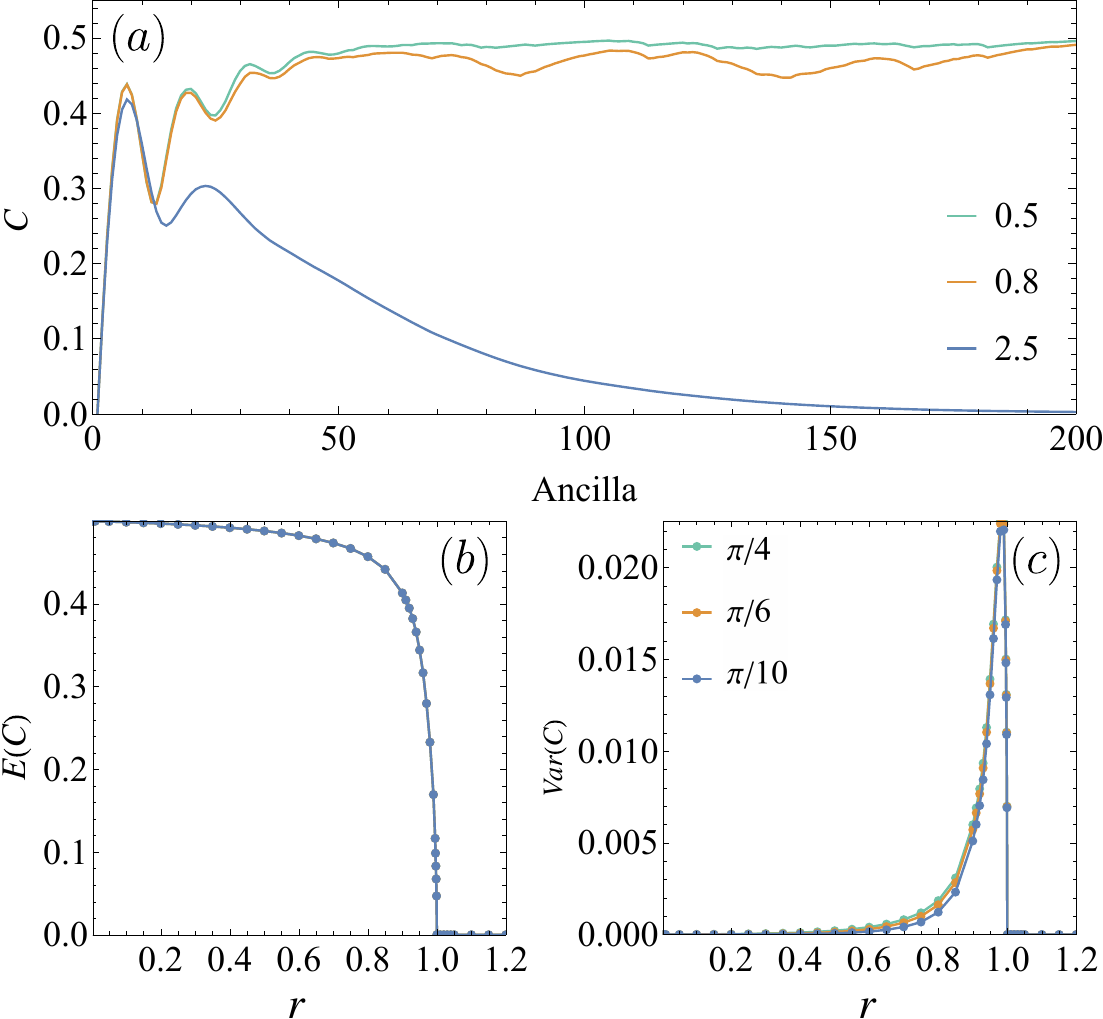}
    \caption{$(a)$ Single shot coherence $\mathcal{C}$ as a function of ancilla number (evaluated at the times each ancilla leaves the process), for different values of $r$, with $\gamma = 0.05$ and $g =\pi/12$. Here we consider only waiting time dynamics. $(b)$ Average coherence $E(\mathcal{C})$ and $(c)$ variance of the coherence $Var(\mathcal{C})$ in the long time limit as a function of $r$, for different values of coupling strength $g$, with $\gamma = 0.05$. 
    }
    \label{fig:dynamics-dephasing-ancilla}
\end{figure}

In the second case, we consider that the dephasing channel describes only the ancilla waiting time dynamics. In Fig.~\ref{fig:dynamics-dephasing-ancilla}(a) we plot the coherence after each ancilla collision for different values of $r$. Here, ancillas lose coherence depending on how long they wait in the queue, but the system does not lose any coherence while it is idle. The coherence transfer, therefore, becomes a competition between the partially dephased ancillas and the remaining coherence of the system, which is acquired through previous partial SWAP interactions. This new kind of competition effect gives rise to a peculiar feature that is observed in Fig.~\ref{fig:dynamics-dephasing-ancilla}(b): the average coherence is independent of the interaction strength $g$. This is an artifact of the very long time behavior. Since the system does not lose any coherence when it is idle, what effectively dictates how much coherence it will have in long times is how strongly dephased the ancillas are. For smaller values of $r$, even if ancillas arrive very rarely, they lose very little coherence, so the partial SWAPs will eventually lead the system to a high coherence state. As the frequency of arrivals increases, the queue starts to pile up and as a consequence, the coherence in the long time limit will be smaller because interactions take place with very strongly dephased ancillas. Once $r > 1$ is reached, a transition takes place and the queue grows indefinitely with every ancilla being completely dephased. As a consequence, the system loses all coherence. The model now homogenizes to the completely dephased ancilla state, which is the identity: 
\[ \lim_{n \to \infty} \rho_S^n = \mathcal{E}_{A}(t\to\infty)[\rho_{A}] = \dfrac{1}{2}\mathbb{I}.\]
Note that this is a steady state of the dynamics because the fluctuations vanish, see Fig.~\ref{fig:dynamics-dephasing-ancilla}(c). 
%This model captures the large $\lambda \tau_{SA}$ behavior (including values below and above the $\lambda \tau_{SA} = 1$ transition, i.e. the ``right'' part of Fig.~\ref{fig:dynamics}(b)) of the general qubit model we presented for the QQCM, where we observe that $\mathcal{C}$ is monotonically decreasing with $\lambda \tau_{SA}$ before reaching $\lambda \tau_{SA} = 1$, and zero after. 

%Those two examples where we turn on only one kind of dynamics at a time illustrate one feature of the full QQCM example we discussed in the main text, namely, that there is value $\lambda \tau_{SA} \in (0,1)$ that maximizes the average coherence. This effect is a direct competition between the two channels acting simultaneously. As we observed in the two examples discussed here, turning off the waiting time dynamics essentially describes the small $\lambda \tau_{SA}$ regime while turning off the idle dynamics effectively describes the large $\lambda \tau_{SA}$ regime. In the former, we see that the coherence transfer is monotonically increasing with $\lambda \tau_{SA}$ whereas in the latter it is monotonically decreasing. The optimal coherence point arises as an inflection point between those two behaviors, which are directly related to the queue. In both examples, the transition is still present at $\lambda \tau_{SA} = 1$, where the average and the variance of the coherence are nonanalytic at this point.

\begin{figure}
    \centering
    \includegraphics[width=0.48\textwidth]{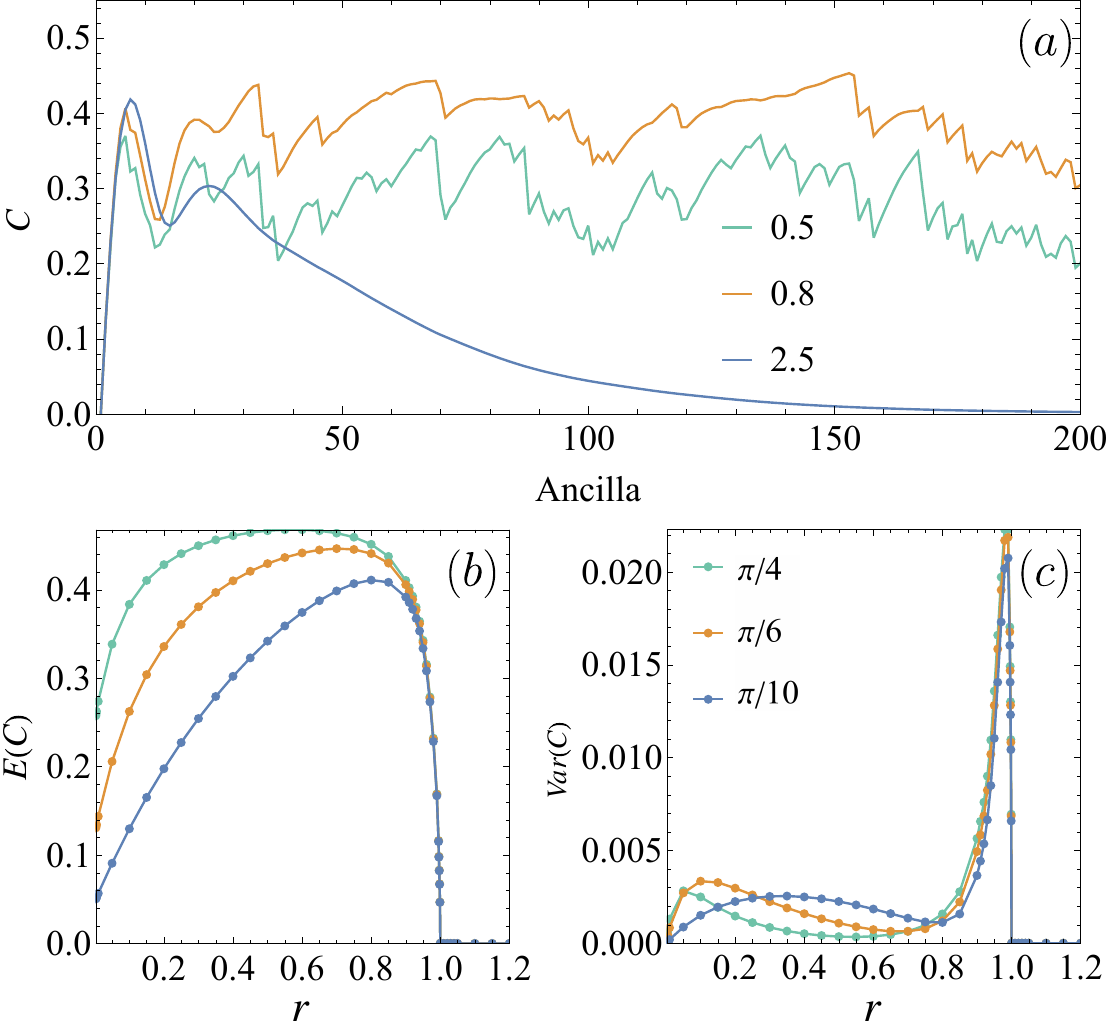}
    \caption{$(a)$ Single shot coherence $\mathcal{C}$ as a function of ancilla number (evaluated at the times each ancilla leaves the process), for different values of $r$, with $\gamma = 0.05$ and $g =\pi/12$. $(b)$ Average coherence $E(\mathcal{C})$ and $(c)$ variance of the coherence $Var(\mathcal{C})$ in the long time limit as a function of $r$, for different values of coupling strength $g$, with $\gamma = 0.05$. 
    }
    \label{fig:dynamics-appendix}
\end{figure}

\textcolor{black}{After we analyzed the effect that each dynamical channel has in the long time dynamics of the system, we now turn our attention to the case where both the waiting time and the idle dynamics are described by the dephasing channel with some $\gamma$.}
In Fig.~\ref{fig:dynamics-appendix}(a) we plot the coherence after each ancilla collision 
for different values of $r$. 
In the first few collisions, the coherence grows.
For the two smaller $r$ curves, it oscillates roughly around a constant, while for $r>1$ it drops down to zero. 
This happens because for $r \gg 1$ the queue quickly becomes unbounded and the ancilla waiting times start to diverge.
A more systematic analysis is shown in Fig.~\ref{fig:dynamics-appendix}(b), where we study average coherence in the long time limit, after very many collisions.
The quantity is plotted as a function of $r$, showing a clear transition at $r = 1$, where $E(\mathcal{C})$ is nonanalytic. %This abrupt transition reflects a known property of the size of G/G/1 queues. 
More interestingly, the coherence is not monotonic with $r$, which reflects the nontrivial interplay of waiting and idle times and, consequently, the competition between the idle/waiting time dynamical channel. 
There is, therefore, an optimal value $r \in [0,1]$, which depends on the quantum dynamics parameters $g$ and $\gamma$, for which the coherence reaches a maximum. 
\textcolor{black}{This feature is due exclusively to the queueing process. The idle dynamics captures the small $r$ behavior (``left'' part of Fig.~\ref{fig:dynamics-appendix}(b)) whereas the waiting time dynamics is the dominant process for $r \sim 1$ (``right'' part of Fig.~\ref{fig:dynamics-appendix}(b)). The competition between the two channels and their interplay (which ultimately boils down to the underlying queueing process) gives rise to the optimal point observed in Fig~\ref{fig:dynamics-appendix}(b).}
%\begin{equation}
%E(\mathcal{C}) = \lim\limits_{n\to \infty} \frac{1}{n}\sum\limits_{j=1}^n \mathcal{C}_j.
%\end{equation}

%In fact, $\mathcal{C}$ is identically zero when $\lambda \tau_{SA} \to 0$ as well. 

Another important feature of the queueing dynamics is that, depending on the queue parameter $r$, the system may never reach a steady state. The dynamics will cause the system state to fluctuate for arbitrarily long times. 
We illustrate this in Fig.~\ref{fig:dynamics-appendix}(c), where we plot the single variance of the coherence $Var(\mathcal{C})$, computed in the long time limit.
%\begin{equation}
%    {\rm var}(\mathcal{C}) = \lim\limits_{n\to \infty} \frac{1}{n-1}\sum_{j=1}^n \big[C_n -E(\mathcal{C})\big]^2.
%\end{equation}
Following what happened to the average, the variance of the coherence is also nonanalytic at $r = 1$.  We see that for $r>1$ the fluctuations tend exactly to zero, which characterizes a true steady state. 
Conversely, for $r < 1$ the fluctuations are nonzero, even in the long time limit. 
This happens because in this regime the queue randomly alternates between idle and busy periods, as depicted in Fig.~\ref{fig:diagrams-and-timeline}(e).

\end{document}